\documentclass[11pt]{article}

\usepackage{array}
\usepackage{cite}
\usepackage[subrefformat=parens]{subcaption}
\usepackage{subdepth}
\usepackage{url}
\usepackage{fullpage} 
\usepackage{graphicx}
\usepackage{authblk}
\usepackage{amsmath}
\usepackage{amsfonts}
\usepackage{algorithm}
\usepackage{algorithmicx}
\usepackage{algpseudocode}

\usepackage[backref, hidelinks]{hyperref}
\usepackage[dvipsnames]{xcolor}

\renewcommand{\arraystretch}{1.25}
\newcommand{\bhline}{\noalign{\hrule height 0.8pt}}

\title{Fast Sign Retrieval via Sub-band Convolution:\\An Elementary Extension of Binary Classification}

\author[1]{Fuma Ito}
\author[1]{Chihiro Tsutake}
\author[1]{Keita Takahashi}
\author[1]{Toshiaki Fujii}

\affil[1]{
Department of Information and Communication Engineering, 
Nagoya University, 
Furo-cho, 
Chikusa-ku, 
Nagoya, 
464-8603, 
Japan}

\date{\empty}

\begin{document}

\maketitle
\vspace{-10mm}

\begin{abstract}
To efficiently compress the sign information of images,
we address a sign retrieval problem for the block-wise discrete cosine transformation~(DCT):
reconstruction of the signs of DCT coefficients from their amplitudes.
To this end, we propose a fast sign retrieval method on the basis of binary classification machine learning.
We first introduce 3D representations of the amplitudes and signs,
where we pack amplitudes/signs belonging to the same frequency band into a 2D slice,
referred to as the sub-band block.
We then retrieve the signs from the 3D amplitudes via binary classification,
where each sign is regarded as a binary label.
We implement a binary classification algorithm using convolutional neural networks,
which are advantageous for efficiently extracting features in the 3D amplitudes.
Experimental results demonstrate that our method achieves accurate sign retrieval with an overwhelmingly low computation cost.
\end{abstract}


\maketitle

\section{Introduction}
\label{s1}
\vspace{-2mm}
Compressing sign information in data is a challenging problem,
and it plays a fundamental role in a wide range of research fields.
In image coding, 
a large number of bits, approximately $20\%$ of the total bits,  
generated by an image encoder are allocated for the signs, as reported in \cite{Sole12}.
Against this background, many previous works~\cite{Tu02,Ponomarenko07,Koyama12,Miroshnichenko18} have developed sign compression methods within the context of image coding,
specifically focusing on compressing signs of discrete cosine transform~(DCT)~\cite{Ahmed74} coefficients.

\begin{figure*}[!t]
\centering
\includegraphics[scale=1.0]{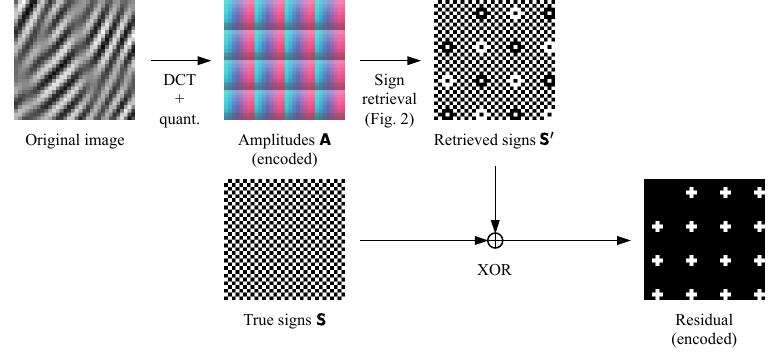}
\vspace{-2mm}
\caption{Encoding process in \cite{Tsutake21}.}
\vspace{4mm}
\label{fig:encdec}
\end{figure*}

To reduce the bits for signs, in our earlier work~\cite{Tsutake21,Suzuki24},
we slightly modified a standardized image encoder and decoder, such as JPEG,
as summarized in Fig.~\ref{fig:encdec}.
In the encoder, DCT coefficients of an image are separated into the signs,
referred to as the \textit{true signs}, and amplitudes; the latter is encoded in the first place.
The encoder locally reconstructs the signs using a \textit{sign retrieval} method whose algorithmic procedure is shared with the decoder.
Instead of encoding the true signs, 
we encode residuals~(XORs) between the true signs and retrieved ones.
In the decoder, an image is decoded by following the reverse path of the encoding process,
where the XORs between the residuals and retrieved signs perfectly reconstruct the true signs.
If the signs are retrieved correctly to some extent, the residuals have many zeros but few ones; 
the residuals can be compressed using entropy coding methods,
e.g., \cite{Huffman52,Rice71,Martin79,Golchin97}.
Therefore, the accuracy of the retrieved signs is crucial for successful sign compression.

\begin{figure*}[!t]
\centering
\begin{subfigure}{1.00\textwidth}
\centering
\includegraphics[scale=1.0]{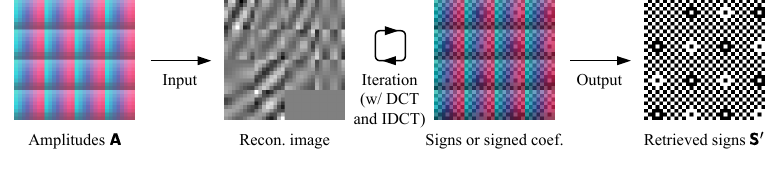}
\caption{Previous methods~\cite{Tsutake21,Suzuki24,Lin24,Sidiropoulos23}}
\vspace{2mm}
\label{fig:sr_a}
\end{subfigure}
\begin{subfigure}{1.00\textwidth}
\centering
\includegraphics[scale=1.0]{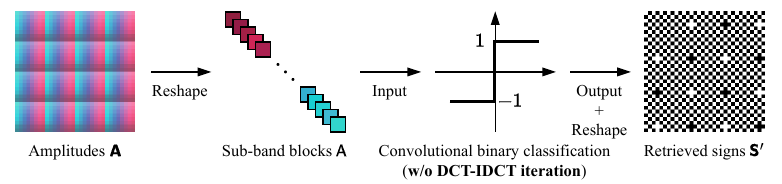}
\caption{Proposed}
\label{fig:sr_b}
\end{subfigure}
\caption{Sign retrieval methods.}
\label{fig:sr}
\end{figure*}

Various sign retrieval methods have been proposed,
as summarized in Fig.~\ref{fig:sr_a}.
To the best of our knowledge, 
all of the previous methods, without exception, are based on iterative optimization.
For example, in \cite{Tsutake21,Suzuki24},
an image and signed DCT coefficients are alternatively retrieved,
and the signs of the retrieved coefficients are extracted.
In \cite{Sidiropoulos23}, 
the signs are regarded as variables and are optimized in an iterative manner.
Previous methods have demonstrated the ability to accurately retrieve a large portion of the signs;
for example, as reported in \cite{Suzuki24}, roughly $75$\% of the total signs can be retrieved.
However, 
these methods come at an excessive computation cost,
primarily due to computing the DCT and the inverse DCT~(IDCT) for each iteration.
For example, as reported in \cite{Suzuki24}, 
it takes approximately 10~s to retrieve $256 \times 256$ signs,
making sign retrieval impractical for real-time encoding and decoding scenarios.

To accelerate the speed of sign retrieval, 
we propose a fast method \textbf{without the DCT-IDCT iteration}, which is summarized in Fig.~\ref{fig:sr_b}.
Unlike previous methods, our method retrieves the signs in a direct manner, 
where sign retrieval is regarded as a variant of a binary classification problem.
We first introduce 3D representations of the amplitudes and signs,
where we pack amplitudes/signs belonging to the same frequency band into a 2D slice,
referred to as the \textit{sub-band block}~\cite{Tu02}.
We then retrieve the signs from the 3D amplitudes via binary classification,
where each sign is regarded as a binary label.
We implement a binary classification algorithm using convolutional neural networks~(CNNs),
which are advantageous for efficiently extracting features in the 3D amplitudes.
Experimental results demonstrate that our method achieves accurate sign retrieval with an overwhelmingly low computation cost:
only \textbf{0.93\%} of the execution time taken for previous methods.
\vspace{2mm}

\noindent\textbf{Limitations:} Throughout this paper,
we utilize constant-size blocks, as was done in JPEG~\cite{Wallace91},
which is not compatible with the variable-size ones in state-of-the-art image coding standards such as
high-efficiency video coding~(HEVC)~\cite{Sullivan12} and versatile video coding~(VVC)\cite{Bross21}.
However, 
we believe that our promising results (discussed in Section~\ref{s4}) will contribute to further advancements in the field of image coding.

\section{Related work}
\label{s2}
\vspace{-2mm}
\subsection{Sign retrieval}
\label{s2ss1}
The sign-retrieval-based method~\cite{Tsutake21}, illustrated in Fig.~\ref{fig:encdec},
is a seminal work in sign compression.
Sign retrieval is known as a special instance of phase retrieval:
namely, determining the signs $\pm1$ is identical to retrieving their phases, $0$ and $\pi$, on the Gaussian plane.
Therefore, sign retrieval has typically been addressed on the basis of phase retrieval approaches.
Tsutake et al.~\cite{Tsutake21} proposed an $\ell_1$-norm minimization method,
motivated by the prior works on phase retrieval~\cite{Bahmani17,Goldstein18}.
Lin et al.~\cite{Lin24} proposed a similar method based on total variation minimization.
Suzuki et al.~\cite{Suzuki24} proposed a CNN-based least squares method.
Sidiropoulos et al.~\cite{Sidiropoulos23} proposed a quadratic programming method.
All of these methods, without exception, retrieve the signs via iterative processes,
as illustrated in Fig.~\ref{fig:sr_a}.
For example,
an image and signed DCT coefficients are alternatively computed in \cite{Tsutake21,Suzuki24},
similar to the Gerchberg-Saxton method~\cite{Gerchberg72} and Fienup method~\cite{Fienup82}.
The Frank-Wolfe method~\cite{Frank56}, utilized in \cite{Sidiropoulos23}, is another example,
which is a variant of a gradient method and computes the DCT to determine the descent direction.
Due to the iterative nature, the DCT and IDCT should be  conducted in each iteration, leading to significant  computation costs.
\vspace{2mm}

\begin{figure*}[!t]
\centering
\includegraphics[scale=1.0]{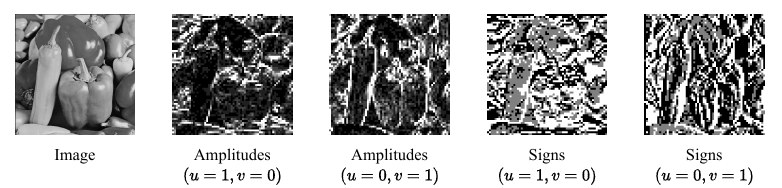}
\vspace{2mm}
\caption{Sub-band blocks obtained from image, $\times 8$ enlarged. Larger amplitudes are represented by brighter luminance. Positive and negative signs are represented by white and black, respectively. Black regions in amplitudes and gray ones in signs correspond to non-significant coefficients.}
\label{fig:corr}
\end{figure*}

\subsection{Sign compression without sign retrieval}
\label{s2ss2}
Ponomarenko et al.~\cite{Ponomarenko07} proposed a prediction-based method similar to the sign retrieval-based method~\cite{Tsutake21},
where the amplitudes are encoded before sign compression.
Their method predicts the signs block-by-block by solving an amplitude-constrained combinatorial problem,
where the signs that minimize an objective function are searched.
To reconstruct the true signs at the decoder,
residuals between the true signs and predicted ones are encoded.
As reported in \cite{Ponomarenko07},
the bit amount for the residuals is $60$--$85$\% of that for the true signs.
Subsequent works, such as \cite{Henry16,Nakagawa17,Filippov19},
have been developed to enhance the accuracy and computation efficiency.

Clare et al.~\cite{Clare11} proposed a sign compression method, known as sign data hiding,
which is a fundamental component in state-of-the-art image coding standards,
HEVC~\cite{Sullivan12} and VVC~\cite{Bross21}. 
This method allows us to skip encoding a single sign of DCT coefficients for each block.
The decoder can compensate for the missing sign by executing a parity check. 
As reported in\cite{Clare11}, 
sign data hiding achieves a BD-rate~\cite{Bjontegaard01}~(a rate-distortion metric) of approximately $-0.6\%$ compared to the HM~4.0 anchor\footnote{
\url{https://hevc.hhi.fraunhofer.de}}.
Sign data hiding has been improved in subsequent works, 
such as \cite{Wang12,Zhang13,Song18}, 
where the encoding of multiple signs per block can be skipped. 

Tu and Tran~\cite{Tu02} proposed a sign compression method inspired by sub-band coding~\cite{Shapiro93,Said96}.
They constructed the so-called sub-band blocks,
which include amplitudes/signs belonging to the same frequency band,
and the true signs are encoded and decoded in the sub-band domain.
Figure~\ref{fig:corr} shows examples of sub-band blocks obtained from an image,
where $u$ and $v$ represent frequency indices along the horizontal and vertical directions, respectively.
As can be seen, each element is strongly correlated with its spatial and sub-band neighbors.
This remarkable statistical feature of the sub-band representation is imported into our method.

\section{Proposed method}
\label{s3}
\vspace{-2mm}
\subsection{Notations}
\label{s3ss1}
A scalar variable and a scalar-valued function are denoted by a regular typeface.
A constant value is denoted by a capital case.
A 3D tensor, 3D-tensor-valued function, and their elements are denoted by a sans-serif typeface.
A 4D tensor and its elements are denoted by a bold sans-serif typeface.
A set is denoted by a Greek letter.

\begin{figure}[!t]
\centering
\begin{subfigure}{0.4\textwidth}
\centering
\includegraphics[scale=1.0]{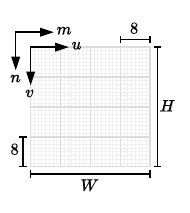}
\vspace{-2mm}
\caption{4D representation}
\label{fig:coor_a}
\end{subfigure}
\begin{subfigure}{0.4\textwidth}
\centering
\includegraphics[scale=1.0]{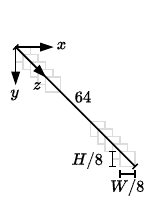}
\vspace{-2mm}
\caption{3D representation}
\label{fig:coor_b}
\end{subfigure}
\caption{Coordinate systems.}
\label{fig:coor}
\end{figure}

As shown in Figs.~\ref{fig:encdec} and \ref{fig:sr}, 
the amplitudes of quantized DCT coefficients are denoted by a 4D tensor $\boldsymbol{\mathsf{A}}$,
and the true and retrieved signs are denoted by $\boldsymbol{\mathsf{S}}$ and $\boldsymbol{\mathsf{S}}'$, respectively.
Figure~\ref{fig:coor_a} illustrates the coordinate system of $\boldsymbol{\mathsf{A}}$, $\boldsymbol{\mathsf{S}}$, and $\boldsymbol{\mathsf{S}}'$.
The width and height of an image are denoted by $W$ and $H$, respectively.
The block size is fixed to $8 \times 8$.
Frequency indices of the DCT basis function are denoted by $u \in [0, 8)$ and $v \in [0, 8)$.
Block indices are denoted by $m \in [0, W/8)$ and $n \in [0, H/8)$.
The $(u,v)$-th element in the $(m,n)$-th block of $\boldsymbol{\mathsf{A}}$, $\boldsymbol{\mathsf{S}}$, and $\boldsymbol{\mathsf{S}}'$ is denoted by $[\cdot]_{u,v,m,n}$.
As shown in Fig.~\ref{fig:coor_b}, we introduce 3D representations of $\boldsymbol{\mathsf{A}}$, $\boldsymbol{\mathsf{S}}$, and $\boldsymbol{\mathsf{S}}'$,
which are denoted by $\mathsf{A}$, $\mathsf{S}$, and $\mathsf{S}'$, respectively.
Their axes are $x$, $y$, and $z$, and the $(x,y,z)$-th element is denoted by $[\cdot]_{x,y,z}$.
\vspace{2mm}

\subsection{Sign retrieval via sub-band convolution}
\label{s3ss2}
To accelerate sign retrieval, we propose a fast method without the DCT-IDCT iteration, 
which is summarized in Fig.~\ref{fig:sr_b}.
Our method retrieves the signs based on binary classification machine learning.
Because processing 4D tensors as they are requires a large computational cost,
we reshape the 4D amplitudes and signs into 3D ones, as follows:
\begin{align}
\mathsf{A}_{x,y,z}&=\boldsymbol{\mathsf{A}}_{u(z),v(z),m(x),n(y)}\label{eq:reshape1}\\
\mathsf{S}_{x,y,z}&=\hspace{0.2mm}\boldsymbol{\mathsf{S}}_{u(z),v(z),m(x),n(y)},
\label{eq:reshape2}
\end{align}
where $u(z)= \lfloor z/8 \rfloor$, $v(z)= z\,\text{mod}\,8$, $m(x)=x$, and $n(y)=y$.
The amplitudes/signs belonging to the $(u(z), v(z))$-th frequency band are packed in the $z$-th 2D slice,
referred to as the sub-band block, and the 3D tensors contain all the $64$ sub-band blocks, 
as shown in Fig.~\ref{fig:corr}.
We simply stack these $64$ sub-bands along the $z$ dimension;
consideration of even/odd symmetry of the DCT kernels might lead to a better method for handling these sub-bands, but we leave it as the future work.
The effectiveness of the reshaping process will be demonstrated in Section~\ref{s4ss2}.

We aim to reconstruct the signs $\mathsf{S}$ from the 3D amplitudes $\mathsf{A}$ via binary classification,
where each sign is regarded as a binary label.
We implement a binary classification algorithm using CNNs,
which have the capability to efficiently extract features in $\mathsf{A}$ along the spatial and sub-band directions.
Table~\ref{tab:network} shows the network architecture of our CNN,
where $I$ represents the number of layers.
The Conv-$i$ has a 3D kernel of the size $3 \times 3 \times C$ ($3 \times 3$: spatial kernel size and $C$: the number of input channels) for each output channel, where all the channels (frequency sub-bands) can interact together in each convolution step.
The spatial kernel size $3 \times 3$ in the sub-band domain corresponds to $24 \times 24$ pixels in the original image domain; thus, a sufficiently large spatial neighbor is considered in each convolution step.
%
%
Our CNN maps the 3D amplitudes $\mathsf{A}$ into a sign tensor of the size $W/8 \times H/8 \times 63$, where the number of channels, 63, corresponds to the AC components.
We denote the inferred signs by $\mathsf{F}(\mathsf{A},\Theta)$, 
where $\Theta$ represent a set of learnable parameters including convolution kernels and biases.

The training procedure is as follows.
Let $\Phi=\{(\mathsf{A}[k],\mathsf{S}[k]):k\in[0,K)\}$ be a set of 3D amplitudes and 3D signs for training,
where $K$ denotes the number of data.
We define the empirical risk
\begin{equation}
r(\Phi,\Theta)=
\frac{1}{WH}\frac{1}{K}\sum_{x,y,z}\sum_{k}
l(\mathsf{F}_{x,y,z}(\mathsf{A}[k],\Theta),\mathsf{S}_{x,y,z}[k]),
\label{eq:emp}
\end{equation}
where $l$ is a loss function.
We obtain the optimal parameter, denoted by $\tilde{\Theta}$, by minimizing the empirical risk:
\begin{equation}
\tilde{\Theta} = \mathop{\mathrm{argmin}}_{\Theta} \,\, r(\Phi,\Theta).
\label{eq:min}
\end{equation}
The solution to \eqref{eq:min} is obtained using the Adam optimizer.

The loss function $l$ in \eqref{eq:emp} is defined as follows.
Let 
\begin{equation}
b(\mathsf{S}_{x,y,z})=
\begin{cases}
1     & \text{if} \quad \mathsf{S}_{x,y,z} = 1\\
0     & \text{otherwise}
\end{cases}
\end{equation}
be a zero-one representation of the true sign $\mathsf{S}_{x,y,z}$.
We utilize the binary cross-entropy function as the loss function: 
\begin{equation}
l(\mathsf{F}_{x,y,z},\mathsf{S}_{x,y,z})=-b(\mathsf{S}_{x,y,z})
\log (\mathsf{F}_{x,y,z})
-(1-b(\mathsf{S}_{x,y,z})) 
\log (1-\mathsf{F}_{x,y,z}),
\label{eq:bce}
\end{equation}
where $(\mathsf{A},\Theta)$ of $\mathsf{F}_{x,y,z}$ has been omitted for notation convenience.
%
%
If the $(x,y,z)$-th amplitude is zero, we set the corresponding loss value to $0$.

At the test phase, 
we obtain the retrieved signs by thresholding $\mathsf{F}_{x,y,z}(\mathsf{A}, \tilde{\Theta})$ as follows.
\begin{equation}
\mathsf{S}_{x,y,z}'=
\begin{cases}
\hspace{0.14mm}+\hspace{0.14mm}1 & \text{if} \quad \mathsf{F}_{x,y,z} \geq 1/2\\
-\hspace{0.015mm}1 & \text{otherwise}
\end{cases}
\end{equation}
Let $x(m)=m$, $y(n)=n$, and $z(u,v)=8u+v$.
A 4D representation of the retrieved signs can be computed as follows.
\begin{equation}
\boldsymbol{\mathsf{S}}_{u,v,m,n}'=\mathsf{S}_{x(m),y(n),z(u,v)}'
\end{equation}
We encode the residuals between the true signs $\boldsymbol{\mathsf{S}}$ and the retrieved ones $\boldsymbol{\mathsf{S}}'$,
as shown in Fig.~\ref{fig:encdec}.

\section{Experimental results}
\label{s4}
\vspace{-2mm}
\subsection{Configuration}
\label{s4ss1}
For training our CNN, 
we randomly cropped $K=\textrm{6,056}$ images with a size of $W \times H = 512 \times 512$ from the CLIC~2020 dataset~\cite{Challenge}. 
The bit-depth was 8~bits/pixel, and each pixel value was normalized within the range 0--1.
We applied block-wise DCT to the images.
We quantized the DCT coefficients using the quality factor~(QF) of 75.
We reshaped the 4D amplitudes and signs in accordance with \eqref{eq:reshape1} and \eqref{eq:reshape2};
as a result, we obtained 3D tensors with a size of $W/8 \times H/8 \times 64=64 \times 64 \times 64$. 
We utilized the Adam optimizer with the learning rate $2 \times 10^{-4}$. 
The batch size and the number of epochs were $256$ and 15,000, respectively. 
We varied the number of convolution layers $I$~(Table~\ref{tab:network}) from $2$ to $8$.
For convenience, we refer to the proposed method trained on the above configuration as the \textit{sub-band convolution method}.

\begin{table}[!t]
\caption{Network architecture.}
\vspace{-6mm}
\label{tab:network}
\begin{center}    
\footnotesize
\renewcommand{\arraystretch}{1.2}
\begin{tabular}{c|wc{14mm}c wc{14mm} c wc{14mm}}
\bhline
Layer       & Conv-$0$             & $\cdots$ & Conv-$i$     & $\cdots$ & Conv-$I$    \\
Ker. size   & $3 \times 3 \times 64$         & $\cdots$ & $3 \times 3\times 128$ & $\cdots$ & $3 \times 3\times 128$\\
In/out ch.  & $64/128$             & $\cdots$ & $128/128$    & $\cdots$ & $128/63$    \\
Act.        & ReLU                 & $\cdots$ & ReLU         & $\cdots$ & Sigmoid     \\
Input       & 3D amp. $\mathsf{A}$ & $\cdots$ & Conv~$i-1$   & $\cdots$ & Conv~$I-1$  \\\bhline
\end{tabular}
\end{center}
\end{table}

\begin{table}[!t]
\caption{Computing environment.}
\vspace{-6mm}
\label{tab:comp}
\begin{center}
\footnotesize
\renewcommand{\arraystretch}{1.2}
\begin{tabular}{wc{36.2mm}|wc{46.2mm}}  \bhline
CPU                     & Intel Core i9-13900KF\\ 
Main memory   & 32~GB \\ 
OS          & Ubuntu 20.04 LTS \\ 
Language \& framework & Python 3.10.11 \& PyTorch 1.13.1\\\bhline
\end{tabular}
\end{center}
\end{table}

To evaluate the effectiveness of the reshaping processes in \eqref{eq:reshape1} and \eqref{eq:reshape2}, 
we employed the following alternative approach, referred to as the \textit{naive method}.
We treated the collection of all amplitudes (or signs) depicted in Fig.~\ref{fig:coor_a} as a plane with dimensions $W \times H = 512 \times 512$.
We reconstructed the sign planes from the amplitude planes,
where we nullified the reshaping process in the sub-band convolution method.
To adapt the network architecture for planes, 
we changed the input channel of Conv-0 and the output channel of Conv-$I$ to $1$.
For fair comparison,
the training process follows the same configuration as that employed in the sub-band convolution method.

\begin{figure*}[!t]
\centering
\hspace{-2.4mm}
\includegraphics[scale=1.01]{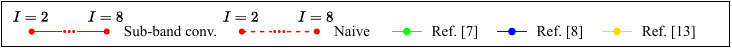}
\begin{subfigure}{0.48\textwidth}
\centering
\includegraphics[scale=0.65]{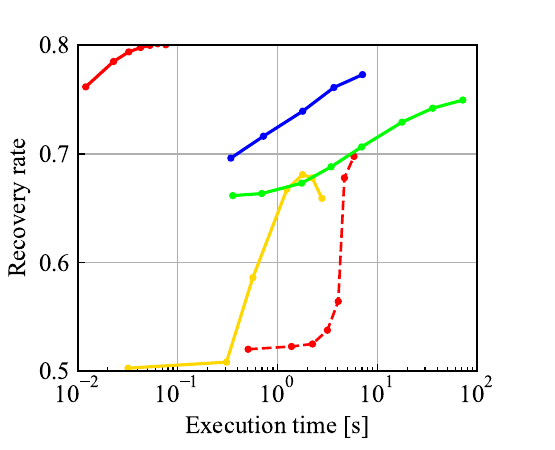}
\vspace{-2mm}
\caption{QF 15}
\end{subfigure}
\begin{subfigure}{0.48\textwidth}
\centering
\includegraphics[scale=0.65]{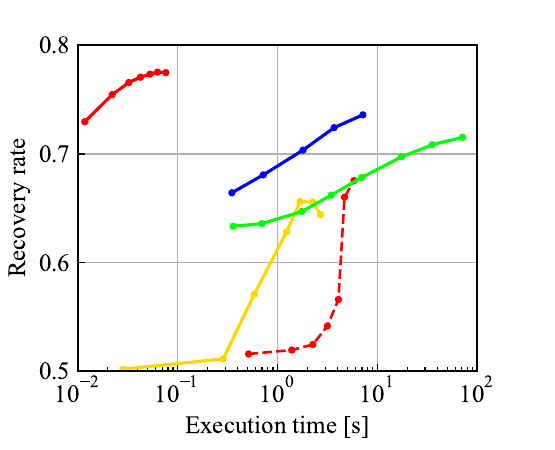}
\vspace{-2mm}
\caption{QF 30}
\end{subfigure}\\
\begin{subfigure}{0.48\textwidth}
\centering
\includegraphics[scale=0.65]{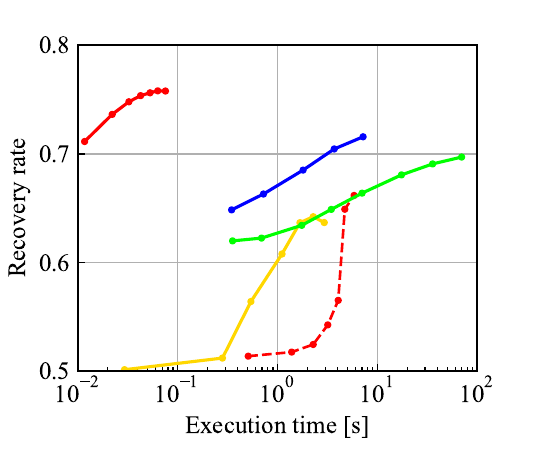}
\vspace{-2mm}
\caption{QF 45}
\end{subfigure}
\begin{subfigure}{0.48\textwidth}
\centering
\includegraphics[scale=0.65]{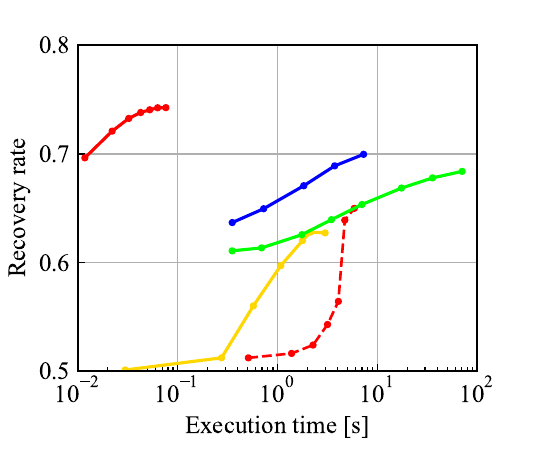}
\vspace{-2mm}
\caption{QF 60}
\end{subfigure}\\
\begin{subfigure}{0.48\textwidth}
\centering
\includegraphics[scale=0.65]{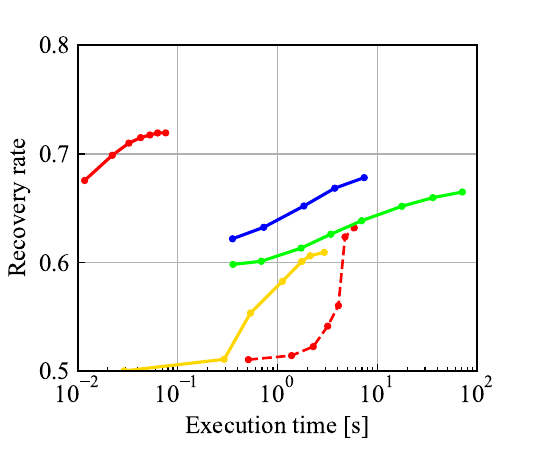}
\vspace{-2mm}
\caption{QF 75}
\label{fig:rr_img_e}
\end{subfigure}
\begin{subfigure}{0.48\textwidth}
\centering
\includegraphics[scale=0.65]{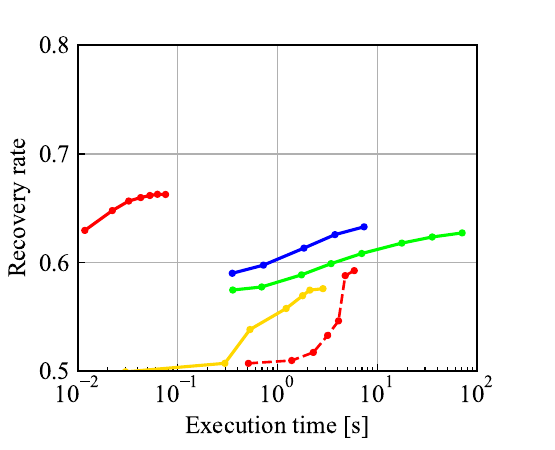}
\vspace{-2mm}
\caption{QF 90}
\end{subfigure}
\caption{Recovery rate against execution time.}
\label{fig:rr_img}
\end{figure*}

\begin{figure}
\centering
\hspace{3.0mm}
\begin{subfigure}{0.45\textwidth}
\hspace{+2.5mm}
\includegraphics[scale=0.73]{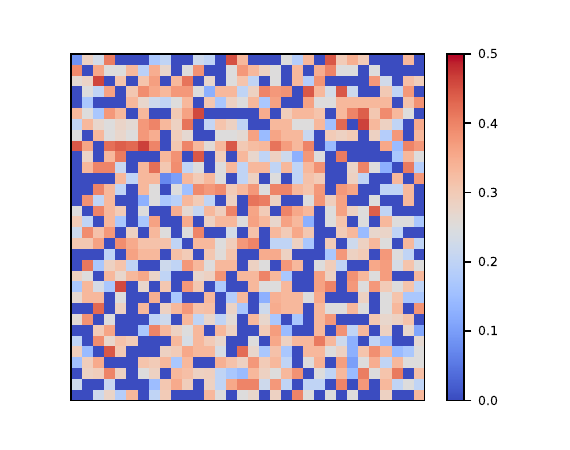}
\vspace{-8mm}
\caption{Worst}
\label{fig:hm_a}
\end{subfigure}
\hspace{3.8mm}
\begin{subfigure}{0.45\textwidth}
\hspace{+2.5mm}
\includegraphics[scale=0.73]{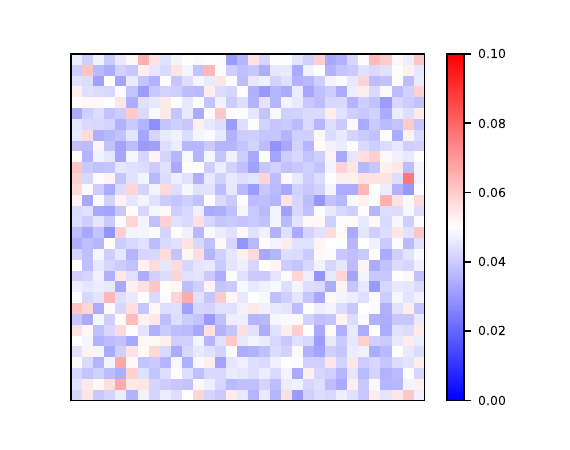}
\vspace{-8mm}
\caption{Variance}
\label{fig:hm_b}
\end{subfigure}
\caption{Heat maps of recovery rates for each block.}
\label{fig:hm}
\end{figure}

For evaluation, we sampled $60$ images from the CLIC~2021 dataset~\cite{Challenge},
which were not included in the training dataset.
The DCT coefficients of the images were quantized with QFs of 15, 30, 45, 60, 75, and 90.
The signs were retrieved using our methods~(sub-band convolution and naive) and previous methods~\cite{Tsutake21,Suzuki24,Lin24};
the latter were iterative methods requiring per-iteration DCT and IDCT (discussed in Section~\ref{s2ss1}).
For the previous methods, we changed the maximum number of iterations from $1$ to $200$.
All the experiments were conducted on the computing environment in Table~\ref{tab:comp};
all the methods were executed using the same CPU.

To quantify the accuracy of the retrieved signs, we define the following recovery rate:
\begin{align}
\frac{\#(\text{correctly retrieved signs of AC coefficients})}{\#(\text{signs of AC coefficients})},\nonumber
\end{align}
where the signs of DC coefficients and non-significant coefficients are excluded.
We also measured the execution time~[s] of all the methods for sign retrieval.
\vspace{2mm}

\subsection{Results}
\label{s4ss2}
Figure~\ref{fig:rr_img} shows recovery rates against execution times,
which were averaged over all blocks in the 60 test images.
The seven points in our method represent results for different values of $I \in [2,8]$,
while the points in the previous methods represent results for different iteration counts.
%
%
For all the QFs,
the sub-band convolution method perfectly outperformed the previous methods in terms of the recovery rate while achieving the shortest execution time.
Specifically, 
the execution time for the maximum recovery rate was $6.74\times 10^{-2}$~s on average, 
i.e., \textbf{0.93\%} of $7.24$~s for the state-of-the-art method~\cite{Suzuki24}.
It is worth noting that the sub-band convolution method also outperformed the naive approach, which demonstrates the effectiveness of the reshaping process.

\noindent\textbf{Remark~1:} 
We report recovery rates for each block to investigate their biases.
We cropped $256 \times 256$ regions from the $60$ test images and fed amplitudes of their DCT coefficients~(QF 75) into our CNN~($I=8$).
We computed the average of recovery rates for each block and obtained the worst value and the variance over the $60$ images.
Figure~\ref{fig:hm} illustrates the results.
We observe that there are clusters of low recovery rates and high variance values at the boundaries.

\vspace{2mm}

\noindent\textbf{Remark~2:} 
For all the methods, the recovery rates tend to increase as QF decreases.
For small QFs, significant DCT coefficients are concentrated at the low frequency bands,
which simplifies sign retrieval.
Therefore, all the methods achieved high recovery rates at small QFs.
In contrast, 
for large QFs, DCT coefficients are distributed across all the frequency bands,
which increases the difficulty of sign retrieval; all the methods thus obtained low recovery rates.
\vspace{2mm}

\section{Conclusion}
\label{s5}
\vspace{-2mm}
In this work,
we addressed a sign retrieval problem for the block-wise discrete cosine transformation~(DCT):
reconstruction of the signs of DCT coefficients from their amplitudes.
To accelerate sign retrieval,
we proposed a fast sign retrieval method without the DCT-IDCT iteration,
where the signs are retrieved based on binary classification machine learning.
We first introduced 3D representations of the amplitudes and signs and then implemented our method by convolutional neural networks,
which are advantageous for extracting features in the 3D amplitudes.
%
%
Our future work will include the extension of our method to the current state-of-the-art image coding standards, 
which utilize variable-size blocks.
We also need to consider other transformations than the DCT, 
e.g., the discrete sine transform and low-frequency non-separable secondary transform~\cite{Koo19}.
We believe that this research direction has the potential to enhance the efficiency of image coding.

\bibliographystyle{IEEEtran}
\bibliography{main}

\end{document}